\documentclass[pre,superscriptaddress,reprint,showpacs,twocolumn,notitlepage,nobalancelastpage]{revtex4-2}

\usepackage{graphicx} 
\usepackage{tikz}
\usepackage{amsmath}
\usepackage{amssymb}
\usepackage{color}
\usepackage{float}
\usepackage{epstopdf}
\usepackage[normalem]{ulem}

\usepackage{pgfplots,pgfplotstable}
\usepackage{soul}
\usepackage{accents}
\usepackage{nameref}
\usepackage{balance}
\usepackage{upgreek}
\usepackage[all]{nowidow}
\usepackage{comment}
\usepackage{dsfont}
\usepackage{cleveref}
\usepackage{array} 				
\usepackage{booktabs} 			
\usepackage{ctable}
\newcolumntype{q}[1]{>{\setlength{\parindent}{0em}}p{#1}}

\crefformat{section}{\S#2#1#3}
\crefformat{subsection}{\S#2#1#3}
\crefformat{subsubsection}{\S#2#1#3}

\makeatletter
\newcommand*{\balancecolsandclearpage}{
  \close@column@grid
  \clearpage
  \twocolumngrid
}
\makeatother

\usetikzlibrary[shadings]
\usetikzlibrary{shapes}
\usetikzlibrary{snakes}
\usepackage{bm}
\usepgfplotslibrary{fillbetween}
\tikzset{
    axis break gap/.initial=10mm
}
\usetikzlibrary{patterns}

\pgfplotsset{compat=newest}

\sethlcolor{white}

\AtBeginDocument{
\heavyrulewidth=.08em
\lightrulewidth=.05em
\cmidrulewidth=.03em
\belowrulesep=.65ex
\belowbottomsep=0pt
\aboverulesep=.4ex
\abovetopsep=0pt
\cmidrulesep=\doublerulesep
\cmidrulekern=.5em
\defaultaddspace=.5em
}

\newcommand{\bn}{\bm{\nabla}}
\newcommand{\bsig}{\bm{\sigma}}
\newcommand{\bu}{\bm{u}}
\newcommand{\bveps}{\bm{\varepsilon}}
\newcommand{\bmu}{\bm{\mu}}
\newcommand{\bD}{\bm{D}}
\newcommand{\bE}{\bm{E}}
\newcommand{\btau}{\bm{\tau}}
\newcommand{\bchi}{\bm{\chi}}
\newcommand{\bpi}{\bm{\pi}}
\newcommand{\bx}{\bm{x}}
\newcommand{\bxch}{\check{\bm{x}}}
\newcommand{\bG}{\bm{G}}

\newcommand{\bC}{\bm{C}}
\newcommand{\bA}{\bm{A}}
\newcommand{\bB}{\bm{B}}
\newcommand{\bSt}{\widetilde{\bm{S}}}
\newcommand{\bWt}{\widetilde{\bm{W}}}

\begin{document}

\title{Generalized effective dynamic constitutive relation for heterogeneous media: \\
Beyond the quasi-infinite and periodic limits}

\author{Jeong-Ho Lee}
\author{Zhizhou Zhang}
\author{Grace X. Gu}
\email[]{ggu@berkeley.edu}
\affiliation{Department of Mechanical Engineering, University of California, Berkeley, CA 94720}

\begin{abstract}
Dynamic homogenization theories are powerful tools for describing and understanding the behavior of heterogeneous media such as composites and metamaterials. However, a major challenge in the dynamic homogenization theory is determining Green's function of these media, which makes it difficult to predict their effective constitutive relations, particularly for the finite-size and/or non-periodic media in real-world applications.  In this paper, we present a formulation for finding the elastodynamic effective constitutive relations for general heterogeneous media, including finite-size and non-periodic ones, by taking into account boundary terms in the Hashin-Shtrikman principle. Our proposed formulation relies on the infinite-body Green's function of a reference homogeneous medium, making it free from the difficulty of determining Green's function even for the homogenization of finite-size media. Additionally, we demonstrate the universal applicability of this formulation for both random and deterministic heterogeneous media. This work contributes to a better understanding of the homogenization theory and the design of next-generation metamaterials that require the accurate prediction of effective material characteristics for dynamic wave manipulation under desired operating environments.
\end{abstract}

\maketitle

\section{introduction}
Heterogeneous media are considered a significant class of engineering materials, where their microstructures can be controlled to adapt to new operating environments and exhibit superior properties compared to raw materials. Metamaterials, a type of heterogeneous media consisting of building blocks large enough to be fabricated yet small enough to affect macroscopic material properties, have opened up new opportunities in the field of advanced materials. They exhibit unprecedented characteristics, particularly in manipulating dynamic material responses such as negative-index, subwavelength focusing, and invisibility cloaking for electromagnetic waves {\cite{smith2000negative,pendry2000negative,schurig2006metamaterial,pendry2006controlling,leonhardt2009broadband}}, collimators, superlenses, and inaudible cloaks for elastic (acoustic) waves {\cite{fang2006ultrasonic,ambati2007surface,zhang2009focusing,zhang2011broadband,quan2015mimicking,cummer2016controlling,quan2018maximum,lee2022maximum}}. Extensive experimental and theoretical work has been performed to explore the rational designs of tailored metamaterials with desired macroscopic properties {\cite{engheta2006metamaterials,liu2009broadband,wilt2020accelerating, kadic2015experiments,shaltout2019spatiotemporal,chen2020nano}}.

Dynamic homogenization theories are powerful tools to describe and understand the behavior of heterogeneous media, by seeking to define effective constitutive relations and material characteristics that govern macroscopic wave propagation of a hypothetical homogenized medium. Willis first presented a linear elastodynamic homogenization theory based on the ensemble average technique and Green's function of heterogeneous media {\cite{willis1980polarization1,willis1980polarization2,willis1981variational1,willis1981variational2,willis1983overall,willis1985nonlocal,Willis1997,willis2011effective,willis2012construction,willis2019statics}}, from which it is discovered that the homogenized medium can exhibit non-classical features: 1) non-local material properties both in space and time; 2) tensorial density, i.e., not scalar; and 3) non-classical constitutive relation. In this non-classical constitutive relation, a coupling effect exists between linear momentum $\bmu$ and stress $\bsig$ with velocity $\bm{v}$ and strain $\bveps$, called \textit{Willis coupling}, and it has been extended in the case of piezoelectric heterogeneous media to a coupling effect between $\bmu$, $\bsig$, and electric displacement $\bD$ with $\bm{v}$, $\bveps$, and electric field $\bE$, called \textit{electro-momentum coupling} {\cite{pernas2020symmetry,muhafra2022homogenization,zhang2022rational,kosta2022maximizing}}. Remarkably, these coupling effects occur even in the homogenization of deterministic media {\cite{shuvalov2011effective,pernas2021electromomentum}}. In addition, this idea of Willis was particularized for periodic heterogeneous media based on Bloch-Floquet waves {\cite{nemat2011overall,nemat2011homogenization,srivastava2015elastic,nassar2015willis,nassar2016generalized}}.

According to Willis' book chapter {\cite{Willis1997}}, determining Green's function of heterogeneous media is the biggest challenge when it comes to the dynamic homogenization theory. As Willis noted, \textit{``If the Green's function could be determined, then all problems would be solved! Of course, this cannot be done.''} To address this challenge, researchers have used the Hashin-Shtrikman principle {\cite{willis1981variational1,willis1983overall,Willis1997,willis2012construction,willis2019statics,milton2007modifications}}, which involves using Green's function of a reference comparison medium that is homogeneous. However, previous works in the literature have constraints when determining Green's function of the reference medium. For instance, Green's function must satisfy certain boundary conditions, such as being zero along the boundaries (i.e., homogeneous boundary condition) or subject to the given types of boundary conditions of the target heterogeneous medium to be homogenized. Therefore, these approaches can only be suitable for infinite or quasi-infinite heterogeneous media large enough for boundaries to be not important. This limitation along with the Bloch-Floquet wave approach that only works for periodic cases poses challenges when it comes to predicting the effective constitutive relations of finite-size non-periodic heterogeneous media for real-world applications.

This paper presents two novel contributions to the linear elastodynamic homogenization of general heterogeneous media, including finite-size and non-periodic ones, for their effective constitutive relations. First, we introduce a new homogenization formulation based on the Hashin-Shtrikman principle, which takes into account boundary terms that have not been adequately considered previously. This formulation allows the use of the infinite-body Green's function of the reference homogeneous medium. The effective constitutive relations obtained from this formulation display the non-classical features of the homogenized medium. Second, we demonstrate the universal applicability of the proposed formulation for both random and deterministic heterogeneous media. With these two contributions, this work provides a better understanding of the homogenization theory and enables the design of next-generation composites and metamaterials in which securing their effective characteristics is critical for dynamic wave manipulation. 

\section{theory}
Consider a boundary-value problem of a heterogeneous medium $\Omega$ which has random or deterministic microstructures but its boundary conditions and body force are taken to be sure, in arbitrary space dimension.  Here, we proceed with piezoelectric media, which possess both Willis and electro-momentum coupling, as the homogenization formulation of standard (non-piezoelectric) media only with Willis coupling can be easily reduced by taking zero piezoelectricity as shown later in this paper.  Thus, our goal is the development of a general formulation to construct effective dynamic constitutive relations of the heterogeneous medium, i.e.,
\begin{equation}
\label{constitutive_eff}
\left(\begin{matrix} \langle\bm{\sigma}\rangle \\ \langle\bm{D}\rangle \\ \langle\bm{\mu}\rangle \end{matrix}\right) = \left[\begin{matrix} \widetilde{\bm{C}} & \widetilde{\bm{B}}^{T} & \widetilde{\bm{S}}_{1} \\ \widetilde{\bm{B}} & -\widetilde{\bm{A}} & \widetilde{\bm{W}}_{1} \\ \widetilde{\bm{S}}_{2} & \widetilde{\bm{W}}_{2} & \widetilde{\bm{\rho}} \end{matrix}\right]\left(\begin{matrix} :\langle\bveps\rangle \\ \cdot\langle\bn\phi\rangle \\ \cdot\langle\bm{v}\rangle \end{matrix}\right)
\end{equation}
with the transpose operator $(\cdot)^{T}$, in which the quantities with angle brackets mean the homogenized macroscopic ones where $\bveps=\frac{1}{2}\left[\bn\bu+(\bn\bu)^{T}\right]$ with spatial gradient operator $\bn$ and displacement $\bu$, and $\bm{v} = \dot{\bu}$ with over-bar being time derivative, and $\bn\phi = -\bE$ for electric potential $\phi$. The over-tilde is used to denote the effective material properties for stiffness $\bm{C}$, dielectricity $\bm{A}$, and piezoelectricity $\bm{B}$, and mass density $\rho$.  We note here the effective mass density has a tensorial form. $\widetilde{\bm{S}}$ and $\widetilde{\bm{W}}$ are the Willis and electro-momentum coupling terms, respectively.  As shown later in this paper, the effective material properties are non-local operators in general such that, for instance, $\left(\widetilde{\bC}:\langle\bveps\rangle\right)(\bx) = \int{\widetilde{\bC}(\bx,\bxch):\langle\bveps(\bxch)\rangle}~d\Omega(\bxch)$ with respect to the spatial coordinates $\bx$ and $\bxch$.

The local mechanical and electric fields of the heterogeneous medium must satisfy the governing equations on $\Omega$
\begin{equation}
\label{GE}
\begin{cases}
\bn\cdot\bsig + \bm{f} = \dot{\bmu} \\
\bn\cdot\bD = q
\end{cases}
\end{equation}
where $\bm{f}$ is body force, and $q$ is free charge density. The constitutive relations between the local kinetic and kinematic fields are written as
\begin{equation}
\label{constitutive}
\left(\begin{matrix} \bm{\sigma}\\ \bm{D}\\ \bm{\mu} \end{matrix}\right) = \left[\begin{matrix} \bm{C} & \bm{B}^{T} & 0 \\ \bm{B} & -\bm{A} & 0 \\ 0 & 0 & \rho \end{matrix}\right]\left(\begin{matrix} :\bveps \\ \cdot\bn\phi \\ \cdot\bm{v} \end{matrix}\right)
\end{equation}
in which the material properties depend on the microstructure materials of the heterogeneous medium at the local position.  Here, we assume the standard tensor symmetries on the properties $\bm{C}$, $\bm{A}$, and $\bm{B}$. Boundary conditions along the boundary $\partial\Omega$ are given as
\begin{equation}
\begin{cases}
\bsig\cdot\bm{n}=\bm{t} \text{~on~}\partial\Omega_{t};~~~~~~~ \bu=\bu_{bc} \text{~on~}\partial\Omega_{u} \\
\bD\cdot\bm{n}=-c \text{~on~}\partial\Omega_{c};~~~~ \phi=\phi_{bc} \text{~on~}\partial\Omega_{\phi}
\end{cases}
\end{equation}
where $\bm{t}$ and $\bu_{bc}$ are prescribed traction and displacement, respectively, with the outward unit normal vector $\bm{n}$, and $c$ and $\phi_{bc}$ are prescribed surface charge density and electric potential, respectively.  Note that $\partial\Omega_{t}\cap\partial\Omega_{u} = \partial\Omega_{c}\cap\partial\Omega_{\phi} = \varnothing$.

\begin{figure*}
\begin{center}
\includegraphics[width=1.3\columnwidth]{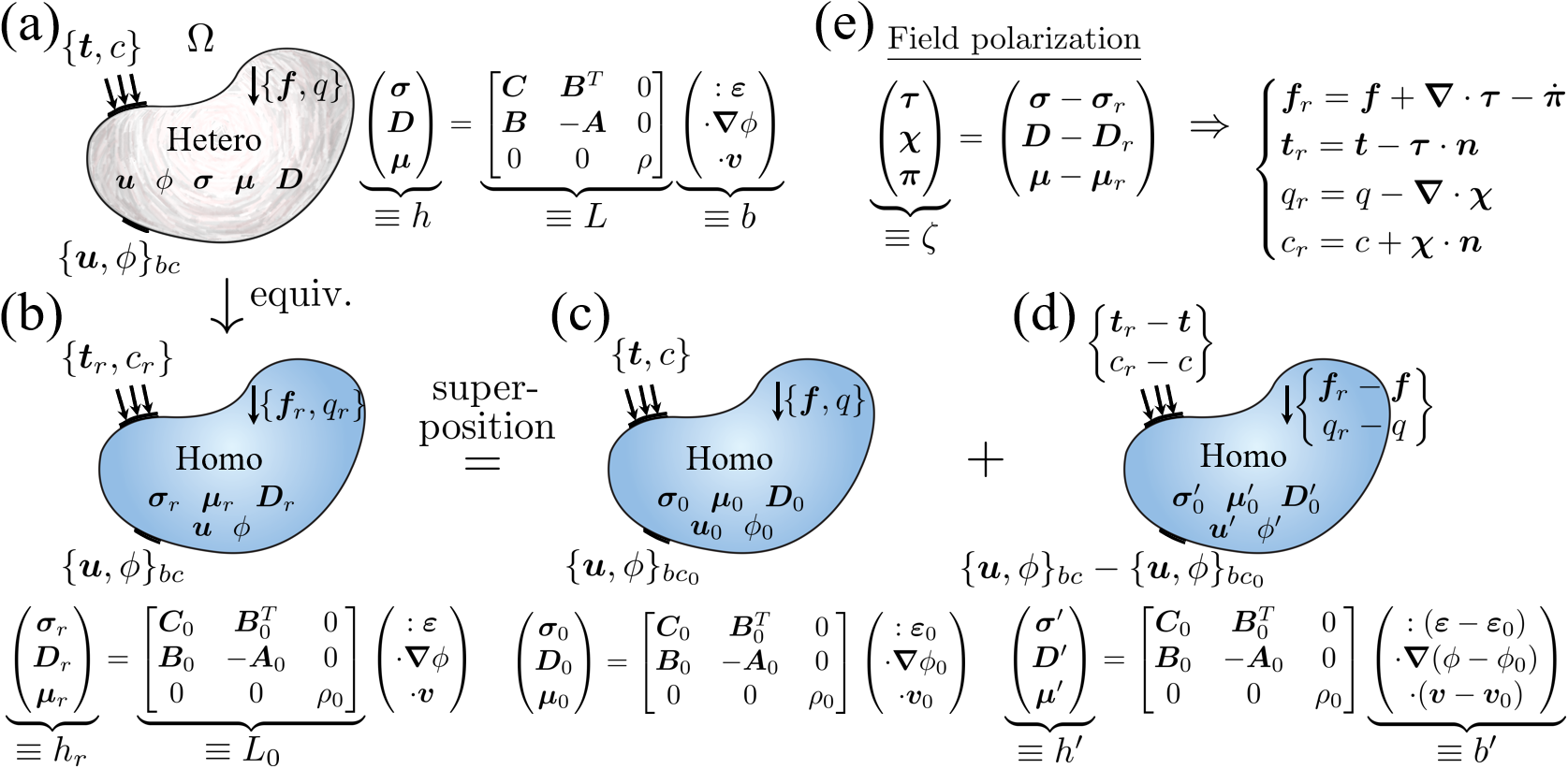}
\vspace{-0mm}
\caption{\label{fig1} Schematics of the Hashin–Shtrikman principle-based approach to equivalently represent a heterogeneous medium (a) via a homogeneous reference medium (b), with the field polarization and its consequent source and boundary terms (e), which is decomposed into two homogeneous systems (c) and (d).
} 
\vspace{-0mm}
\end{center}
\end{figure*}

Now following the Hashin-Shtrikman principle {\cite{hashin1963variational}} schematically shown in Fig.~\ref{fig1}, let's introduce a homogeneous reference medium with constant properties $\bm{C}_{0}$, $\bm{A}_{0}$, $\bm{B}_{0}$, and $\rho_{0}$, which may or may not be one of the microstructure materials of the heterogeneous medium.  Then, field polarizations are defined as 
\begin{equation}
\label{fp}
\left(\begin{matrix} \btau \\ \bchi \\ \bpi \end{matrix}\right) = \left(\left[\begin{matrix} \bm{C} & \bm{B}^{T} & 0 \\ \bm{B} & -\bm{A} & 0 \\ 0 & 0 & \rho \end{matrix}\right] - \left[\begin{matrix} \bm{C}_{0} & \bm{B}_{0}^{T} & 0 \\ \bm{B}_{0} & -\bm{A}_{0} & 0 \\ 0 & 0 & \rho_{0} \end{matrix}\right]\right)\left(\begin{matrix} :\bveps \\ \cdot\bm{\nabla}\phi \\ \cdot\bm{v} \end{matrix}\right)
\end{equation}
where $\btau$, $\bchi$, $\bpi$ are the polarization of stress, electric displacement, and linear momentum, respectively.  Hereafter, we will write matrix equations for the shake of simplicity via symbolic expression based on the notations in Fig.~\ref{fig1} --- for instance of (\ref{constitutive}) and (\ref{fp}), $h=Lb$ and $\zeta = \Delta L b$ with $\Delta L=L-L_0$, respectively.  Then, the boundary-value problem of the heterogeneous medium can be equivalently represented, as shown in Fig.~\ref{fig1}(b), via a boundary-value problem of the homogeneous reference medium with the constant properties $L_{0}$ as
\begin{equation}
\begin{cases}
\bn\cdot\bsig_{r}+(\bm{f}+\bn\cdot\btau-\dot{\bpi}) = \dot{\bmu}_{r} \\
\bn\cdot\bD_{r} = (q-\bn\cdot\bchi)
\end{cases}
\end{equation}
where $\bsig_{r} = \bsig-\btau$, $\bmu_{r} = \bmu-\bpi$, and $\bD_{r} = \bD-\bchi$ such that $h_{r} = L_{0}b$ via the symbolic expression, with boundary conditions
\begin{equation}
\begin{cases}
\bsig_{r}\cdot\bm{n}=\bm{t}-\btau\cdot\bm{n} \text{~on~}\partial\Omega_{t};~~~~~~~ \bu=\bu_{bc} \text{~on~}\partial\Omega_{u} \\
\bD_{r}\cdot\bm{n}=-c-\bchi\cdot\bm{n} \text{~on~}\partial\Omega_{c};~~~~ \phi=\phi_{bc} \text{~on~}\partial\Omega_{\phi}.
\end{cases}
\end{equation}

As we consider a linear elastodynamic system, this equivalent boundary-value problem can be decomposed via the superposition principle as shown in Fig.~\ref{fig1}(c) and (d), in which Fig.~\ref{fig1}(c) is the reference homogeneous medium subjected to the given Neumann boundary conditions of the heterogeneous medium to be homogenized. This leads to a resultant boundary-value problem building on the system of Fig.~\ref{fig1}(d) that is independent of both the source terms $\bm{f}$ and $q$ and the Neumann boundary conditions $\bm{t}$ and $c$ as 
\begin{equation}
\begin{cases}
\bn\cdot\bsig'+(\bn\cdot\btau-\dot{\bpi}) = \dot{\bmu}' \\
\bn\cdot\bD' = (-\bn\cdot\bchi)
\end{cases}
\end{equation}
such that $h' = L_{0}b'$ as the symbolic expression, with boundary condition
\begin{equation}
\begin{cases}
\bsig'\cdot\bm{n}=-\btau\cdot\bm{n} \text{~on~}\partial\Omega_{t};~~~~ \bu=\bu_{bc}-\bu_{bc_0} \text{~on~}\partial\Omega_{u} \\
\bD'\cdot\bm{n}=-\bchi\cdot\bm{n} \text{~on~}\partial\Omega_{c};~~~ \phi=\phi_{bc}-\phi_{bc_0} \text{~on~}\partial\Omega_{\phi}
\end{cases}
\end{equation}
where $\bu_{bc_0}$ and $\phi_{bc_0}$ are arbitrary Dirichlet boundary conditions on the system of Fig.~\ref{fig1}(c).  Then, the problem can be solved via the boundary-element method using the infinite-body Green's function of the reference homogeneous medium.  By employing the Laplace transform to replace the time derivative by the Laplace variable $s = -i\omega$ with angular frequency $\omega$, the solution is obtained as
\begin{widetext}
\begin{equation}
\label{sol}
\left(\begin{matrix} \bu(\bx) \\ \phi(\bx) \end{matrix}\right) \equiv w(\bx) = \langle w(\bx) \rangle + \int{ \Big(\check{B}G(\bx,\bxch)\Big)^{T}\langle\zeta(\bxch)\rangle - \Big(\check{B}G(\bx,\bxch)\Big)^{T}\zeta(\bxch) }~d\Omega(\bxch)
\end{equation}
\end{widetext}
as the symbolic expression with
\begin{equation}
\check{B} = \left[\begin{matrix}
\check{\bn} & 0 \\ 0 & \check{\bn} \\ s & 0
\end{matrix}\right];~~ G(\bx,\bxch) = \left[\begin{matrix}
\bm{G}_{11}(\bx,\bxch) & \bm{G}_{12}(\bx,\bxch) \\ \bm{G}_{21}(\bx,\bxch) & G_{22}(\bx,\bxch)
\end{matrix}\right]
\end{equation}
in which the time-reduced Green's function matrix $G(\bx,\bxch)$ is defined to solve {\cite{norris1994dynamic,wang20053}}
\begin{equation}
\begin{cases}
\begin{aligned}
\bn\cdot\left( \bn\bG_{11}^{T}:\bC_{0}^{T} + \bn\bG_{21}\cdot\bB_{0} \right)-s^{2}\rho_{0}\bG_{11}^{T} &= -\bm{\delta}_{x\check{x}} \\
\bn\cdot\left( \bn\bG_{11}^{T}:\bB_{0}^{T} - \bn\bG_{21}\cdot\bA_{0}^{T} \right) &= \bm{0} \\
\bn\cdot\left( \bn\bG_{12}:\bC_{0}^{T} + \bn G_{22}\cdot\bB_{0} \right)-s^{2}\rho_{0}\bG_{12} &= \bm{0} \\
\bn\cdot\left( \bn\bG_{12}:\bB_{0}^{T} - \bn G_{22}\cdot\bA_{0}^{T} \right) &= -\delta_{x\check{x}} \\
\end{aligned}
\end{cases}
\end{equation}
where $\delta_{x\check{x}}$ is the Dirac-delta function $\delta(\bx-\bxch)$, and $\bm{\delta}_{x\check{x}} = \delta(\bx-\bxch)\bm{I}$ with the identity tensor $\bm{I}$.  We note that the over-check of $\check{B}$ means the gradients are associated with $\bxch$ which comes from the non-local nature of Green's function.  

The solution (\ref{sol}) requires only the infinite-body Green's function of a homogeneous medium, and its integration volume extends over the region occupied by the target heterogeneous medium to be homogenized. While different homogenization theories were derived based on the reference homogeneous medium in the previous literature {\cite{willis1981variational1,willis1983overall,Willis1997,willis2012construction,willis2019statics,milton2007modifications}}, the boundary terms in the Hashin-Shtrikman principle-based approach were not considered before. This results in constraints on Green's function of the reference medium in their homogenization. Therefore, the previous formulations only work for heterogeneous media of infinite (at least quasi-infinite) size or simple geometries to which Green's function can be determined.

The solution (\ref{sol}) yields the kinematic fields as
\begin{equation}
\label{sol2}
b(\bx) = \langle b(\bx) \rangle + \int{ \Gamma(\bx,\bxch)\langle\zeta(\bxch)\rangle - \Gamma(\bx,\bxch)\zeta(\bxch) }~d\Omega(\bxch)
\end{equation}
with $\Gamma(\bx,\bxch) = B\Big(\check{B}G(\bx,\bxch)\Big)^{T}$ where $B$ is replacing the gradients in $\check{B}$ to be associated with $\bx$.  We note that $\Gamma(\bx,\bxch)$ is dominated by its singularity near $\bx=\bxch$ such that it can be written as $\Gamma(\bx,\bxch) = \mathcal{D}\delta_{x\check{x}} + \Gamma^{H}(\bx,\bxch)$. $\mathcal{D}$ is constant which represents the contribution of an infinitesimal region around the singularity position while $\Gamma^{H}(\bx,\bxch)$ is of the contribution outside of the singular region.  This integral equation can be written via a compact integral operator form as $b = \langle b \rangle + \Gamma\langle\zeta\rangle - \Gamma\zeta$.  From this, one can find a relation between $\langle\zeta\rangle$ and $\langle b \rangle$ that can be expressed in a general form of
\begin{equation}
\langle\zeta\rangle = \Gamma^{+}\left[ \Big\langle(\mathcal{I}+\Gamma\Delta L)^{+}\Big\rangle^{+} - \mathcal{I} \right]\langle b \rangle
\end{equation}
in which $\mathcal{I}$ is the identity operator, and $(\cdot)^{+}$ denotes the inverse integral operator, e.g., $\Gamma^{+}\Gamma = \mathcal{I}$.  Therefore, by the definition of field polarizations (i.e., $h=L_{0}b+\zeta$), the effective material property matrix $L_{eff}$, for the effective constitutive relations $\langle h \rangle = L_{eff}\langle b \rangle$ as the compact integral operator form of (\ref{constitutive_eff}), can be finally obtained as
\begin{equation}
\label{L_eff}
L_{eff} = L_{0} + \Gamma^{+}\left[ \Big\langle(\mathcal{I}+\Gamma\Delta L)^{+}\Big\rangle^{+} - \mathcal{I} \right]
\end{equation}
which well exhibits the non-local feature of the effective constitutive relations of the homogenized medium.  This equation can be solved by various methods such as the iteration methods, e.g., the Neumann series and the resolvent method {\cite{LEWIS2022349}}, and the discretization by quadrature rules {\cite{baker1977numerical}} that can provide a direct connection to the finite element method.

This derived formulation is general but when $\bC^{T}=\bC$ (i.e. $C_{ijkl}=C_{lkji}$), $L_{eff}$ is self-adjoint based on the self-adjointness of Green's function, which delivers $\bSt_{2} = \bSt_{1}^{\dagger}$ and $\bWt_{2} = \bWt_{1}^{\dagger}$ with the formal adjoint operator $(\cdot)^{\dagger}$ with respect to the spatial variable, e.g., $\bSt_{1}^{\dagger}(\bx,\bxch) = \bSt_{1}(\bxch,\bx)$ and $\bWt_{1}^{\dagger}(\bx,\bxch) = \bWt_{1}(\bxch,\bx)$.  As (\ref{L_eff}) is derived from (\ref{sol}), it is unique regardless of both source terms and boundary conditions on the heterogeneous medium to be homogenized.  This uniqueness is a reminiscence of the source-driven homogenization {\cite{sieck2017origins}}, and this implies that the non-uniqueness issue in the absence of both source terms and boundary conditions can be mitigated via applying eigenstrains {\cite{willis2011effective}}.  Furthermore, since we haven't made any assumptions on the homogenization strategy to map microscopic quantities to homogenized macroscopic ones, the proposed homogenization formulation is universal.  We note that this formulation approach can also be employed to obtain effective constitutive relations in different physics, e.g., electromagnetism.

\section{validation}
To validate the proposed homogenization formulation and demonstrate its universal applicability, we compare the effective constitution obtained from our formulation to 1-D analytical solutions of problems for random and deterministic heterogeneous media.  

\begin{figure*}
\begin{center}
\includegraphics[width=1\columnwidth]{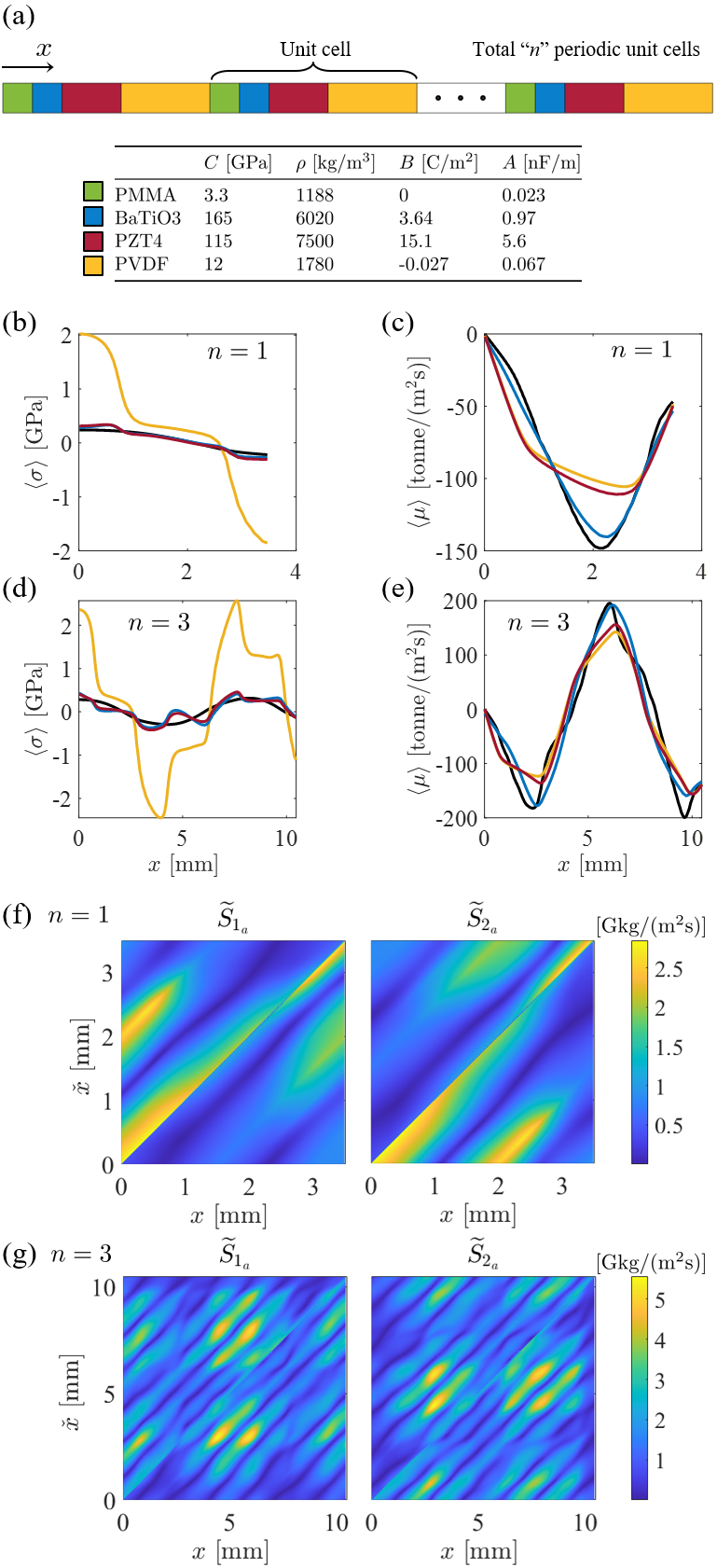}
\vspace{-0mm}
\caption{\label{fig2} (a) Schematics of the 1-D finite-size heterogeneous medium with the $n$ repeating unit cell. (b-e) Comparison of the kinetic fields.  The black line is the ensemble average of analytical solutions of each ensemble.  The blue line denotes macroscopic fields using our formulation whereas the red line is the ones not considering the coupling terms.  The yellow line is a homogenization using the arithmetic mean of material properties. (f and g) Visualization of the apparent coupling coefficients $\widetilde{S}_{1_a}$ and $\widetilde{S}_{2_a}$.
} 
\vspace{-0mm}
\end{center} 
\end{figure*}

First, consider a homogenization problem based on the ensemble average technique for a finite-size random heterogeneous medium with a periodically repeating unit cell, as shown in Fig.~\ref{fig2}(a).  This unit cell consists of PMMA--BaTiO$_3$--PZT4--PVDF layers with 0.5 mm, 0.5 mm, 1 mm, and 1.5 mm of each layer thickness, a total 3.5 mm of the unit cell thickness $l_{uc}$.  Relying on the periodic microstructure geometry, the position of period can be considered as a uniformly distributed random variable that has $l_{uc}^{-1}$ of uniform probability density over the unit cell domain $\Omega_{uc}$.  Therefore, this feature reduces the ensemble average of any $l_{uc}$-periodic function $\Upsilon$ in the homogenization to a spatial average as $\langle\Upsilon(x)\rangle = l_{uc}^{-1}\int_{\Omega_{uc}}{\Upsilon_{y_0}(x-y)}~dy$ in which the $l_{uc}$-periodic function in realization $y\in\Omega_{uc}$ is represented using $\Upsilon_{y_0}(x-y)$ with the subscript $y_0$ denoting the realization at $y=0$.  Assuming the absence of free charge and external electromagnetic fields, the electric governing equation $\nabla D=0$ in (\ref{GE}) gives zero electric displacement across the medium domain, which leads to $\nabla\phi = (BA^{-1})\varepsilon$.  Here, the bold is omitted in a 1-D manner.  In this case, the boundary-value problem of the piezoelectric medium becomes a pure mechanical problem in which the local and effective constitutive relations are modified as
\begin{equation}
\label{constitutive_a}
\left(\begin{matrix} \sigma \\ \mu \end{matrix}\right) = \left[\begin{matrix} C_{a} & 0 \\ 0 & \rho \end{matrix}\right]\left(\begin{matrix} \varepsilon \\ {v} \end{matrix}\right); ~~~
\left(\begin{matrix} \langle\sigma\rangle \\ \langle\mu\rangle \end{matrix}\right) = \left[\begin{matrix} \widetilde{C}_{a} & \widetilde{S}_{1_a} \\ \widetilde{S}_{2_a} & \widetilde{\rho}_{a} \end{matrix}\right]\left(\begin{matrix} \langle\varepsilon\rangle \\ \langle
 v \rangle \end{matrix}\right)
\end{equation}
where the subscript $a$ denotes the apparent quantities as $C_{a} = C + B^{2}A^{-1}$ for the microscopic property, and $\widetilde{C}_{a} = \widetilde{C} + \widetilde{B}^{2}\widetilde{A}^{-1}$, $\widetilde{S}_{1_a} = \widetilde{S}_{1} + \widetilde{B}\widetilde{W}_{1}\widetilde{A}^{-1}$, $\widetilde{S}_{2_a} = \widetilde{S}_{2} + \widetilde{B}\widetilde{W}_{2}\widetilde{A}^{-1}$, and $\widetilde{\rho}_{a} = \widetilde{\rho} + \widetilde{W}_{1}\widetilde{W}_{2}\widetilde{A}^{-1}$ for the effective properties.  By the symbolic expression, we write (\ref{constitutive_a}) as $h_{a}=L_{a}b_{a}$ and $\langle h_{a} \rangle = L_{eff_a}\langle b_{a} \rangle$ respectively for the local and effective relation, with $h_{a} = (\sigma~ \mu)^T$ and $b_{a} = (\varepsilon~ v)^T$.  Then, (\ref{sol2}) is reduced, via the compact integral operator form, as 
\begin{equation}
\label{sol3}
b_{a} = \langle b_{a} \rangle + \Gamma_{a}\langle\zeta_{a}\rangle - \Gamma_{a}\zeta_{a}
\end{equation}
where $\zeta_{a} = (\tau~ \pi)^{T}$, and $\Gamma_{a}(x,\check{x}) = B_{a}\Big(\check{B}_{a}^{T}G_{a}(x,\check{x})\Big)$ with $\check{B}_{a} = (\check{\nabla}~ s)^{T}$ and $B_{a}$ replacing the gradient to be associated with $x$, which can be decomposed as $\Gamma_{a}(x,\check{x}) = \mathcal{D}_{a}\delta_{x\check{x}} +\Gamma_{a}^{H}(x,\check{x})$, where $\mathcal{D}_{a}$ is constant,  based on its singularity near $x=\check{x}$.  The apparent infinite-body time-reduced Green's function $G_{a}(x,\check{x})$ is defined to solve
\begin{equation}
\nabla\left( \nabla G_{a}C_{0_a} \right)-s^{2}\rho_{0}G_{a} = -\delta(x-\check{x})
\end{equation}
where $C_{0_a}$ is the apparent stiffness of the reference homogeneous medium, i.e., {\cite{duffy2015green}}
\begin{equation}
G_{a}(x,\check{x}) = (2ik_{0_a}C_{0_a})^{-1}e^{-ik_{0_a}\vert{x-\check{x}}\vert}
\end{equation}
with $k_{0_a} = \omega\sqrt{\rho_0/C_{0_a}}$.  Then, this leads to
\begin{equation}
\label{L_eff_a}
L_{eff_a} = L_{0_a} + \Gamma_{a}^{+}\left[ \Big\langle(\mathcal{I}+\Gamma_{a}\Delta L_{a})^{+}\Big\rangle^{+} - \mathcal{I} \right]
\end{equation}
where $\Delta L_{a} = L_{a} - L_{0_a}$ with $L_{0_a}$ being the apparent material property matrix of the reference homogeneous medium.  We note that taking the piezoelectricity $B \rightarrow 0$ exactly recovers the effective constitutive relations of standard heterogeneous media with only Willis coupling.  One can easily expand it to the multidimensional case.  Fig.~\ref{fig2}(b-e) show macroscopic stress and linear momentum fields across the medium domain, under prescribing displacements at its left and right ends respectively by zero and 0.2\% of the medium size, for the different number of the unit cell at time frequency $f_{q}(=\frac{\omega}{2\pi})=0.24$ MHz.  The fields obtained using (\ref{L_eff_a}) are compared with the ensemble average of analytical solutions of each ensemble.  These results show that our proposed homogenization formulation provides an excellent estimate of the effective material properties for finite-size random heterogeneous media and displays well the necessity of the non-classical coupling terms in the effective constitutive relations of the homogenized medium, i.e., Willis and electro-momentum coupling.  Fig.~\ref{fig2}(f) and (g) visualize the apparent coupling coefficients and confirm their non-local feature in space and the adjoint relation between $\widetilde{S}_{1_a}$ and $\widetilde{S}_{2_a}$.

\begin{figure*}
\begin{center}
\includegraphics[width=0.65\columnwidth]{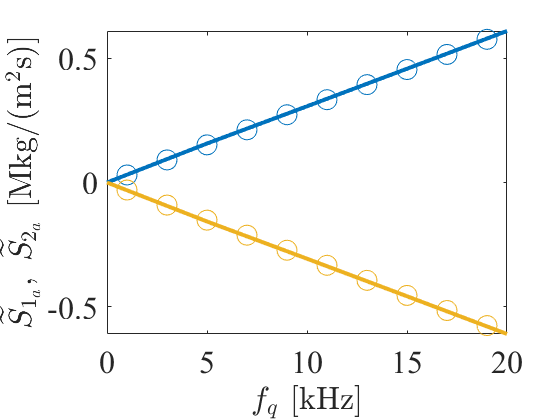}
\vspace{-0mm}
\caption{\label{fig3} The pure imaginary apparent coupling coefficient. The circles are coefficient values obtained from our proposed formulation while the solid lines are the benchmark data of {\cite{pernas2021electromomentum}}. The blue and yellow colors mean the regular and inverted elements, respectively.
} 
\vspace{-0mm}
\end{center} 
\end{figure*}

Second, we extract the non-classical coupling terms of the homogenized medium for a finite-size non-periodic deterministic heterogeneous medium and compare them to benchmark data of {\cite{pernas2021electromomentum}}.  Consider a piezoelectric heterogeneous element made of 2.62 mm of PZT4, 0.19 mm of BaTiO$_3$, and 0.19 mm of PVDF.  We here also assume the absence of free charge and external electric fields.  The microstructures of this element possess wavelength much larger than those characteristic lengths, i.e., $kl \ll 1$ with wavenumber $k$ and length $l$ of the layer.  When a heterogeneous medium is at the subwavelength scale, or when a single isolated element is used rather than a bulk of mutually interacting elements, the effective constitutive relations become local in space, termed \textit{Milton-Briane-Willis equation} {\cite{milton2006cloaking,milton2007new}}.  It has been studied that the local coupling terms originate from broken inversion asymmetry of the microstructures {\cite{muhlestein2016reciprocity,sieck2017origins}}.  Then, the apparent kinematic field vector $b_{a}$ in the modified constitutive relations (\ref{constitutive_a}) can be related to the state vector $\varsigma = (u~ \sigma)^T$ via the material properties as $b_{a} = M\varsigma$ and $\langle b_{a} \rangle = M_{eff}\langle\varsigma\rangle$ via the symbolic expression, with
\begin{equation}
M = \left[\begin{matrix}
0 & C_{a}^{-1} \\ s & 0
\end{matrix}\right]; ~~~
M_{eff} = \left[\begin{matrix}
-s\widetilde{S}_{1_a}\widetilde{C}_{a}^{-1} & \widetilde{C}_{a}^{-1} \\ s & 0
\end{matrix}\right]
\end{equation}
where the derivative of the state vector can also be calculated using the material properties and itself as $\nabla\varsigma = KM\varsigma$ and $\nabla\langle\varsigma\rangle = K_{eff}M_{eff}\langle\varsigma\rangle$, with
\begin{equation}
K = \left[\begin{matrix}
1 & 0 \\ 0 & s\rho
\end{matrix}\right]; ~~~
K_{eff} = \left[\begin{matrix}
1 & 0 \\ s\widetilde{S}_{2_a} & s\widetilde{\rho}_{a}
\end{matrix}\right].
\end{equation}
For the $i$-th layer, the local state vectors at its two ends $x_{L}^{(i)}$ and $x_{R}^{(i)}$ ($x_{L}^{(i)}<x_{R}^{(i)}$) are related via $\varsigma(x_{R}^{(i)}) = T_{l}^{(i)}\varsigma(x_{L}^{(i)})$ with the layer's transfer matrix $T_{l}^{(i)}$ as
\begin{equation}
T_{l}^{(i)} = \left[\begin{matrix}
\cos(k_{a}^{(i)}l^{(i)}) & (C_{a}^{(i)}k_{a}^{(i)})^{-1}\sin(k_{a}^{(i)}l^{(i)}) \\ -C_{a}^{(i)}k_{a}^{(i)}\sin(k_{a}^{(i)}l^{(i)}) & \cos(k_{a}^{(i)}l^{(i)})
\end{matrix}\right]
\end{equation}
with $k_{a}^{(i)} = \omega\sqrt{\rho^{(i)}/C_{a}^{(i)}}$.  By the subwavelength characteristic, $\Gamma_{a}^{H}(x,\check{x})$ can be expressed by expanding it around the element ends upto the linear term such that 
\begin{equation}
\begin{aligned}
\Gamma_{a}^{H}(x,\check{x})
& = \Gamma_{a}^{H}(x-\check{x}) \\
& \approx 
\begin{cases}
\begin{aligned}
\Gamma_{a}^{H}(l_{t}) + \nabla\Gamma_{a}^{H}(l_{t})\left[x-\check{x}-l_{t}\right] ~~~~~~~ &(x\ge\check{x})\\
\Gamma_{a}^{H}(-l_{t}) + \nabla\Gamma_{a}^{H}(-l_{t})\left[x-\check{x}+l_{t}\right] ~~~ &(x<\check{x})
\end{aligned}
\end{cases}
\end{aligned}
\end{equation}
with the element total length $l_{t}$.  Then, based on the continuity of the state vector at the layer intersection and $\langle\zeta_{a}\rangle = (L_{eff_a}-L_{0_a})\langle b_{a} \rangle$, integrating (\ref{sol3}) over $\Omega(x)$ gives an equation to find $L_{eff_a}$ of the deterministic heterogeneous element at the subwavelength scale as 
\begin{equation}
\left[ \sum_{i}^{n_l}\left( \mathcal{E}_{1}^{(i)}+\mathcal{E}_{2}^{(i)}-\mathcal{E}_{3}^{(i)} \right) - (\widetilde{\mathcal{E}}_{1} + \widetilde{\mathcal{E}}_{2} - \widetilde{\mathcal{E}}_{3}) \right]\varsigma(x_{L}^{(1)}) = 0    
\end{equation} 
in terms of the arbitrary $\varsigma(x_{L}^{(1)})$ with $n_l$ being the number of layer, in which the fact that the state vector is sure along the boundary is used, i.e., $\langle\varsigma(x_{L}^{(1)})\rangle = \varsigma(x_{L}^{(1)})$, with
\begin{widetext}
\begin{subequations}
\begin{equation}
\mathcal{E}_{1}^{(i)} = \left(I+\mathcal{D}_{a}\Delta L_{a}^{(i)}\right)K^{(i)^{-1}}\left(T^{(i)}-T^{(i-1)}\right)
\end{equation}
\begin{equation}
\mathcal{E}_{2}^{(i)} = \Gamma_{a-}^{H}\Delta L_{a}^{(i)}\left\{K^{(i)^{-1}}\left(x_{R}^{(i)}T^{(i)}-x_{L}^{(i)}T^{(i-1)}\right) - \left(K^{(i)}M^{(i)}K^{(i)}\right)^{-1}\left(T^{(i)}-T^{(i-1)}\right) \right\}
\end{equation}
\begin{equation}
\mathcal{E}_{3}^{(i)} = \Gamma_{a+}^{H}\Delta L_{a}^{(i)}\left\{K^{(i)^{-1}}\left((x_{R}^{(i)}-l_{t})T^{(i)}-(x_{L}^{(i)}-l_{t})T^{(i-1)}\right) - \left(K^{(i)}M^{(i)}K^{(i)}\right)^{-1}\left(T^{(i)}-T^{(i-1)}\right) \right\}
\end{equation}
\begin{equation}
\widetilde{\mathcal{E}}_{1} = \Big(I+\mathcal{D}_{a}(L_{eff_a}-L_{0_a})\Big)K_{eff}^{-1}\left(T^{(n_l)}-I\right)
\end{equation}
\begin{equation}
\widetilde{\mathcal{E}}_{2} = \Gamma_{a-}^{H}(L_{eff_a}-L_{0_a})\Bigg\{K_{eff}^{-1}\left(l_{t}T^{(n_l)}\right) - \left(K_{eff}M_{eff}K_{eff}\right)^{-1}\left(T^{(n_l)}-I\right) \Bigg\}
\end{equation}
\begin{equation}
\widetilde{\mathcal{E}}_{3} = \Gamma_{a+}^{H}(L_{eff_a}-L_{0_a})\Bigg\{K_{eff}^{-1}\left(l_{t}I\right) - \left(K_{eff}M_{eff}K_{eff}\right)^{-1}\left(T^{(n_l)}-I\right) \Bigg\}
\end{equation}
\end{subequations}
\end{widetext}
where $T^{(i)} = T_{l}^{(i)}T_{l}^{(i-1)} \ldots T_{l}^{(1)}$ with $T_{l}^{(0)} = I$, $\Gamma_{a+}^{H} = \Gamma_{a}^{H}(l_t)$ and $\Gamma_{a-}^{H} = \Gamma_{a}^{H}(-l_t) + \nabla\Gamma_{a}^{H}(-l_t)l_t$, and $I$ is the identity matrix.  This equation can be easily solved by using numerical methods, such as the multivariate Newton--Raphson method, with respect to the components in $L_{eff_a}$, i.e., $\widetilde{C}_{a}$, $\widetilde{S}_{1_a}$, $\widetilde{S}_{2_a}$, and $\widetilde{\rho}_{a}$.  Fig.~\ref{fig3} shows the pure imaginary apparent coupling coefficients of the target heterogeneous element and its inversion, i.e., when the first and third layers are interchanged.  The coupling coefficients obtained from our formulation have $\widetilde{S}_{1_a} = \widetilde{S}_{2_a}$ and are opposite in sign for the regular and inverted elements, i.e., a direction-dependent response, which is coincident with the observation in {\cite{pernas2021electromomentum}}.  The obtained coupling coefficient values are also well agreed with the benchmark data.  These results demonstrate the universal applicability of our proposed formulation not only for ensemble average-based homogenization on random heterogeneous media but also for deterministic cases.

\section{conclusion}
In conclusion, we have derived a formulation to find the dynamic effective constitutive relations for general heterogeneous media, including finite-size and non-periodic ones, by carefully taking into account boundary terms in the Hashin-Shtrikman principle. Our formulation relies on the infinite-body Green's function of a reference homogeneous medium, making it free from the difficulty of determining Green's function even for the homogenization of finite-size media. We note that as our theory generalizes the homogenization theory on two-phase disordered heterogeneous media {\cite{torquato1997effective,torquato1997exact,kim2020effective}}, our formulation satisfies rapid convergence at least up to two-phase disordered heterogeneous media, with the decomposition of $\Gamma(\bx,\bxch)$. It is envisioned that this study will enable more accurate predictions of the effective material properties of heterogeneous media and expand to more real-world applications.

\begin{acknowledgments}
This research was supported by the Defense Advanced Research Projects Agency (Fund Number: W911NF2110363) and Alfred P. Sloan Foundation.
\end{acknowledgments}

\vspace{+0mm}

\vspace{+0mm}

\vspace{+0mm}

\end{document}